\documentclass[a4paper]{jpconf}
\usepackage{graphicx}
\usepackage{longtable}
\usepackage{float}
\usepackage{rotating}
\usepackage{amssymb}
\usepackage{latexsym}
\usepackage{epsfig}
\usepackage{color}

\newcommand{\pu} {\mbox{$p_T$}}
\newcommand{\gjet} {\mbox{$\gamma+jet$}}
\newcommand{\pp} {\mbox{$p+p$}}
\newcommand{\AuAu} {\mbox{$Au+Au$}}

\newcommand{\kt} {\mbox{$k_{T}$}}
\newcommand{\piz} {\mbox{$\pi^0$}}
\newcommand{\rg} {\mbox{$R_{\gamma}$}}
\newcommand{\xe} {\mbox{$x_{E}$}}

\begin{document}
\title{High \pu\ Direct Photon-Hadron Correlations Using the PHENIX Detector}

\author{Matthew Nguyen}

\address{Stony Brook University, Stony Brook, NY 11794 USA}

\ead{manguyen@bnl.gov}

\begin{abstract}
Jet tomography, the study of differential energy loss of hard
scattered partons to infer the density profile of the medium, is
greatly improved by precise knowledge of the initial energy of the
hard probe. As photons are not strongly interacting, the momentum of
the recoil jet from a direct photon trigger is balanced, to a good
approximation, by the momentum of the photon. The energy loss of the
away-side jet may be viewed as an effective modification of the
fragmentation function.   Direct photon-hadron correlations in
$A$+$A$ collisions should be sensitive to modified jet fragmentation
as well as to medium response effects.  Complementary measurements
from p+p collisions are necessary to benchmark jet fragmentation
expectations at $\sqrt{s} = 200 GeV$ as well as to constrain
perturbative calculations in the \gjet\ channel. Here we present new
results from \pp\ and \AuAu\ collisions which use a statistical
method to subtract the background from decay photons.

\end{abstract}

\section{Introduction}

\gjet\ events have long been considered to be a useful probe of
partonic energy loss in nuclear collisions.\cite{wang}  To first
approximation \gjet\ is well represented by the LO QCD Compton
scattering process $q + g \rightarrow q + \gamma$.  Due to the
dominance of this process the \gjet\ cross-section should provide a
constraint on the gluon distribution function in elementary particle
collisions.\cite{owens} In central nuclear collisions at RHIC
energies the hard-scattered parton may lose energy as it traverses
the nuclear medium but the photon will not. In principle, jet energy
loss measurements should constrain the parameter $\hat{q}$ which
which controls the strength of the energy loss. Although di-jet
events occur at a much higher rate than \gjet\ events, the initial
energy of the outgoing partons is hard to determine, complicating
their interpretation.  In the case of prompt photons arising from
the Compton process the initial energy of the jet may be estimated
by measuring the energy of the photon.

Two particle correlation measurements have a history in high energy
physics pre-dating the advent of sophisticated jet reconstruction
algorithms.   These techniques have proven useful with regard to the
high multiplicity RHIC data where the application of these
algorithms is problematic.    Single and di-hadron measurements have
shown the strong suppression of high \pu\ hadron yields
\cite{ppg026} and the disappearance of the away-side jet at high \pu
\cite{starcorr} \cite{ppg083}. However these observables may be
dominated by jet production near the surface of the collision zone
\cite{surface}. Direct photon-hadron correlations should not suffer
from the same geometrical bias. Insofar as the jet kinematics may be
considered fixed by the photon energy, particle production on the
away-side should represent a path length averaged energy loss.
Departure from vacuum fragmentation expectations for away-side
hadron distributions may be viewed as the combined effect of energy
loss and the response of the medium to the deposited energy.

As a matter of nomenclature we consider direct photons to be all
photons not from hadron decay.  Hence we include  photons from jet
fragmentation, which are allowed at NLO. Typically the latter are
removed by isolation requirements, but such cuts are difficult to
apply in heavy-ion collisions. Baseline measurements for
fragmentation photons are necessary as the parton $\rightarrow$
photon fragmentation functions are not well constrained.  Although
the energy loss mechanism should suppress production of these
photons in central \AuAu collisions, additional photon
bremsstrahlung may be induced by the presence of the strong color
fields, which would be an interesting energy loss observable in its
own right. \cite{zak}

\section{Analysis Technique and Results}

The main challenge experimentally is to subtract the large
background associated with photons from meson decay.  In \pp\
collisions direct photons are sub-dominant to decay photons until
approximately
$10 GeV$.  %(comment about $\eta-h$ measurements?).
  In central \AuAu\ collisions direct photons gain greater
significance relative to the decay photon background due to the suppression of high \pu\ jet-associated hadrons.

The decay photon associated yield is deduced from the measure \piz-h associated yield via a Monte Carlo
procedure which takes in to account the decay kinematics and detector effects.
%(Another note on eta-h?).

One may then perform a statistical subtraction of per-trigger yields
using

\begin{equation}
Y_{direct} = \frac{1}{{R_\gamma}-1}\cdot(R_{\gamma}Y_{inclusive
}-Y_{decay}).
\end{equation}

where $Y$ is defined as $dN_{pairs}/N_{trig}$ yield and \rg\ is the
ratio of the number of inclusive photons to the number of decay
photons. \rg\ has been measured in \AuAu\ collisions and may be
derived from the measured direct photon and \piz\ spectra in \pp.

%\subsection{\pp}

\begin{figure}[h]
\begin{center}
\begin{minipage}{14pc}
\includegraphics[width=14pc]{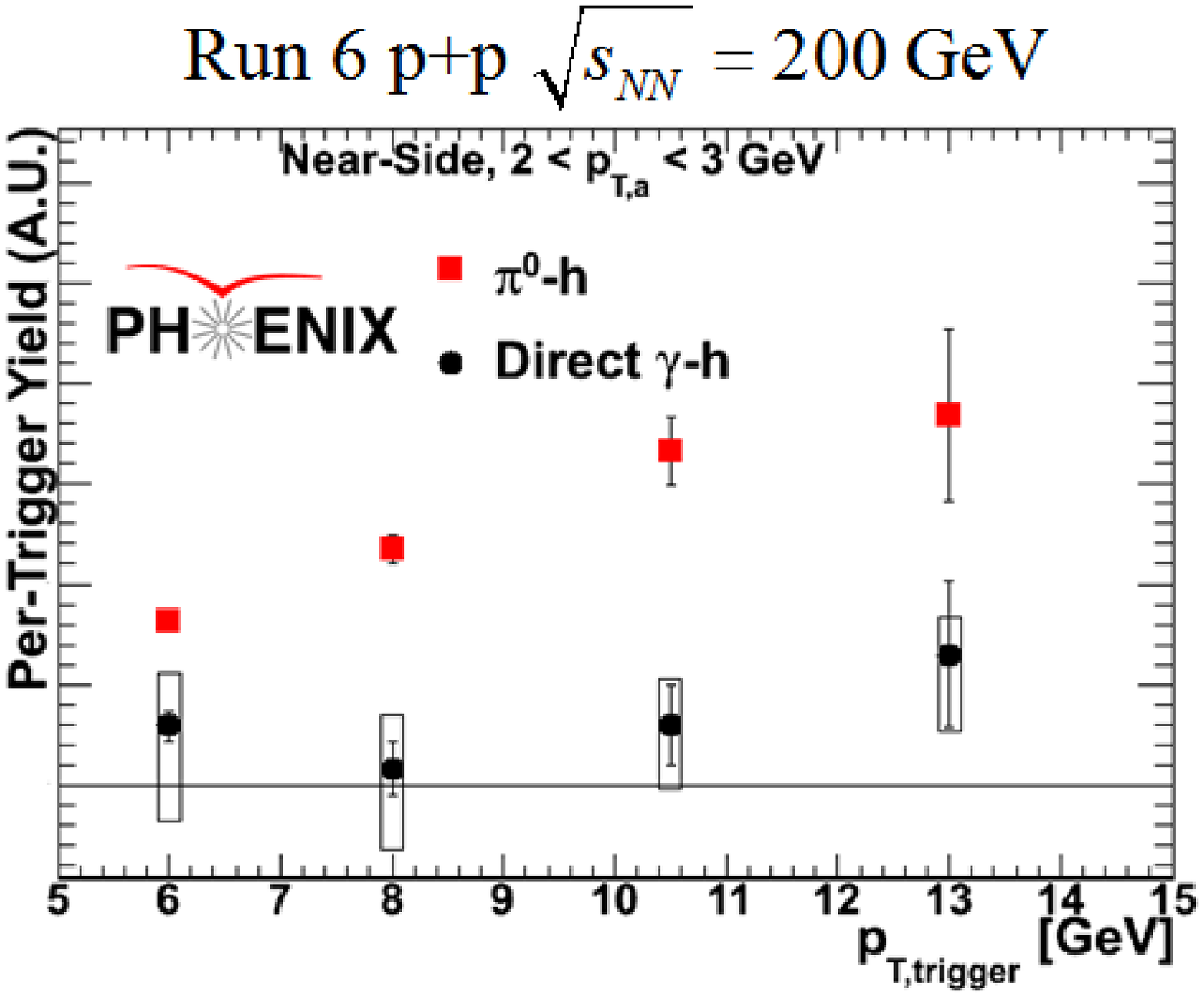}
\caption{\label{ns}  Near-side per-trigger yield of charged hadrons
vs trigger \pu\ for \piz\ triggers ($\textcolor{red}{\fullsquare}$)
and direct photon triggers ($\fullcircle$).}
\end{minipage}\hspace{2pc}%
\begin{minipage}{14pc}
\includegraphics[width=14pc]{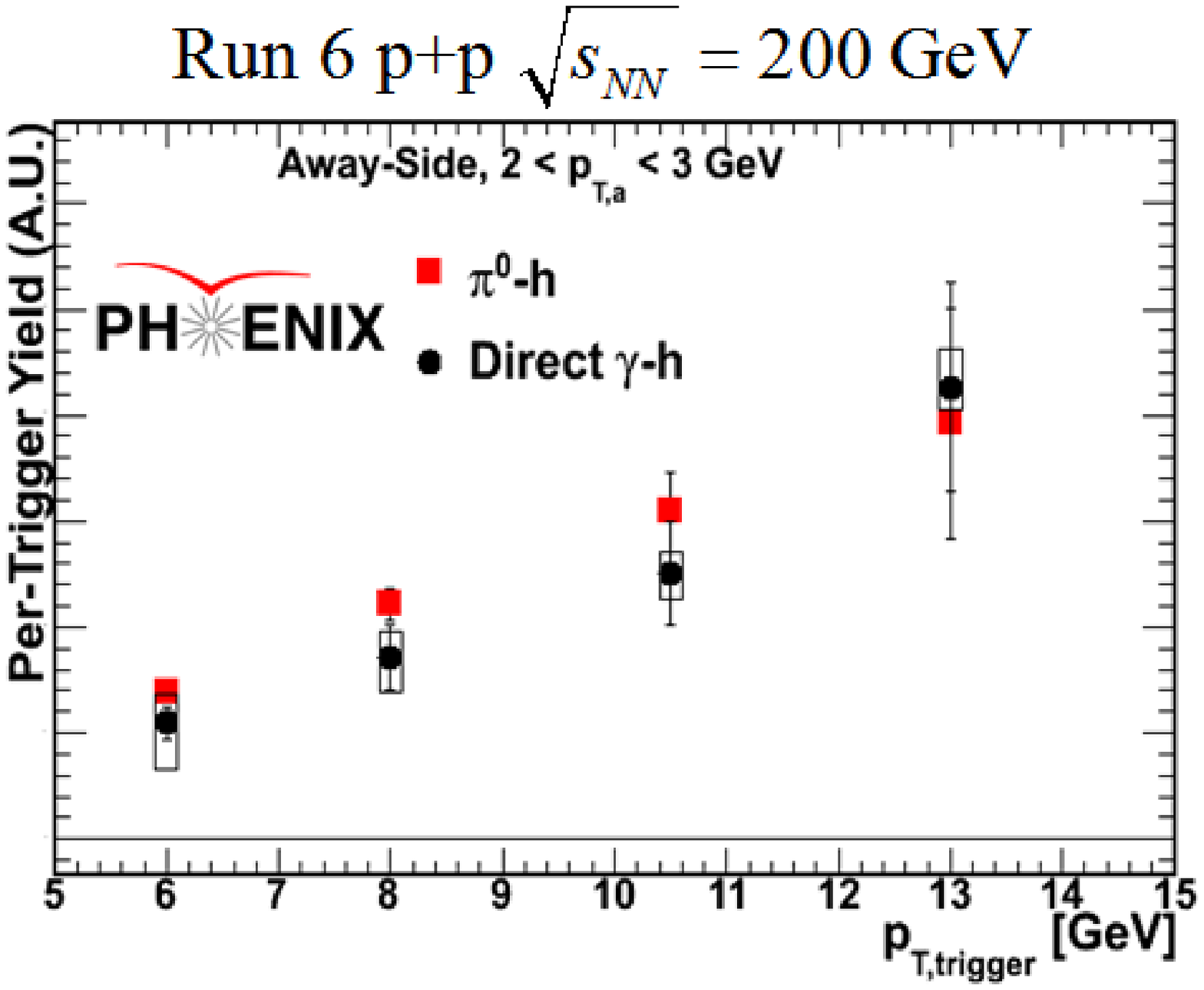} %\\ [2cm]
\caption{\label{as}  Away-side per-trigger yield of charged hadrons
vs trigger \pu\ for \piz\ triggers ($\textcolor{red}{\fullsquare}$)
and direct photon triggers ($\fullcircle$).}
\end{minipage}
\end{center}
\end{figure}

Figures \ref{ns} and \ref{as} show per-trigger yields of charged
hadrons for \piz\ and direct photon triggers from the Run 6 \pp\
data set for the near and away-side, respectively. The systematic
error is dominated by the uncertainty in the \rg\ and the
contribution from heavy meson decay. The measurement extends over a
large kinematic range, from $5 < p_{T,\gamma} < 15$. On the
near-side the data don't allow large particle production from
fragmentation. This is consistent with prompt photon production,
although current uncertainties leave room for some contribution from
near-side fragmentation.  On the away-side the per-trigger yields
for direct photon look similar to \piz\ yields. At fixed trigger
\pu\ The $<Q^2>$ for direct photons should be smaller than for
\piz\'s, resulting in a slightly smaller associated yield.  This
expectation is compatible with the data.

If the contribution from fragmentation photons is not too large
$p_{T,\gamma} \approx p_{T,jet}$.  The distribution of the quantity
\xe\, defined as $-\vec{p}_{T,trigger}\cdot
\vec{p}_{T,associated}/|p_{T,trigger}|^2$ may then be taken as a
good proxy for the fragmentation function of the away-side jet and
should therefore approximately scale with \pu. In \cite{ppg029}
PHENIX showed that this is not the case for \piz\ triggers due the
near-side fragmentation.  Non-scaling effects such as near-side
fragmentation and the \kt\ effect should be well-understood from
\pp\ data and included in any theoretical description of
photon-hadron correlations.  In nuclear collisions this would allow
energy loss effects to be understood quantitatively and
unambiguously.
%footnote on kT effect

\begin{figure}[h]
\begin{center}
\begin{minipage}{14pc}
\includegraphics[width=14pc]{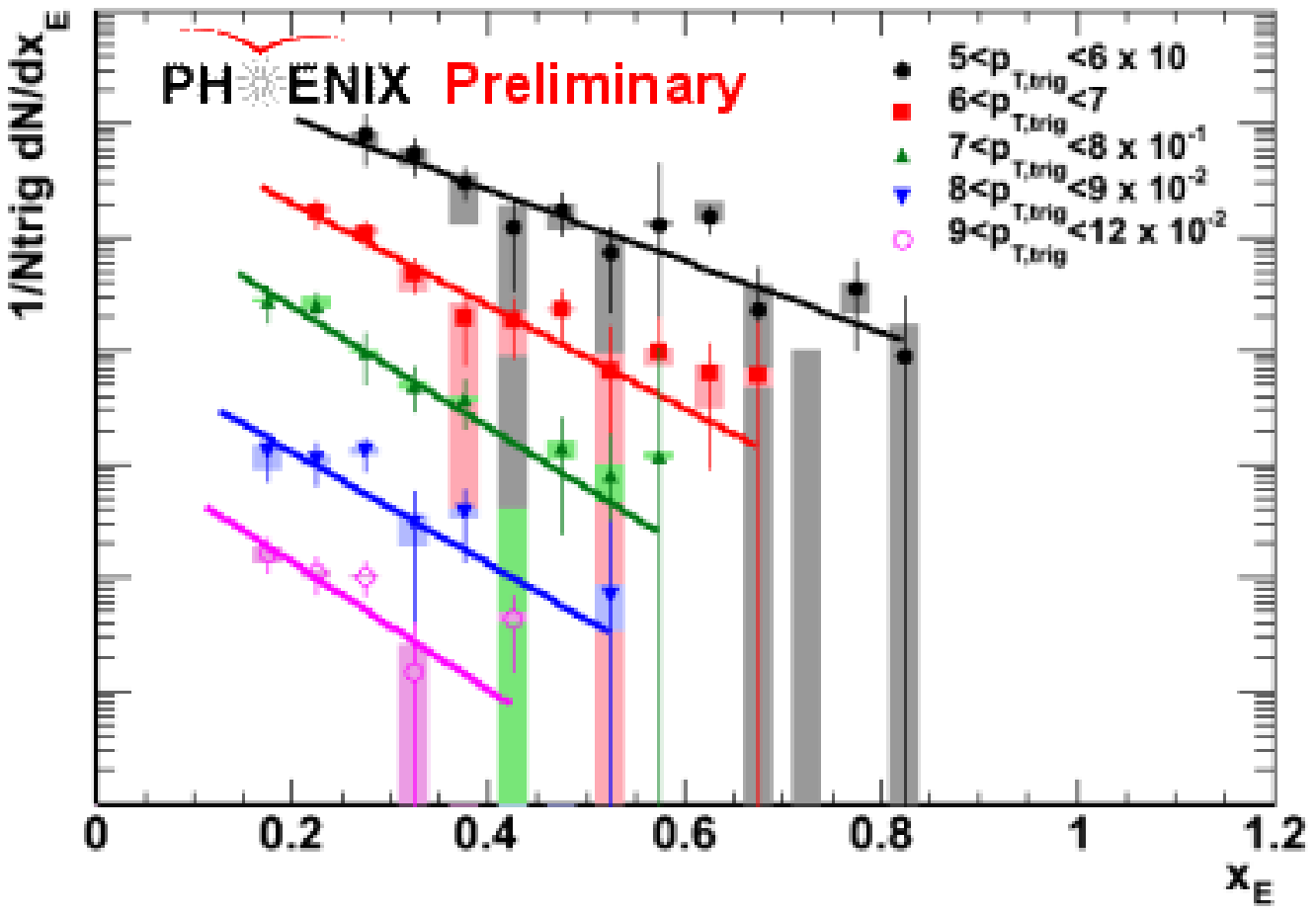}
\caption{\label{xe}\xe\ distributions of direct photon-hadron pairs.
Different $p_{T,\gamma}$ selections are offset by factors of 10 for
clarity.  The \pu\ range for charged hadrons is 1-5 GeV. }
\end{minipage}\hspace{2pc}%
\begin{minipage}{14pc}
\includegraphics[width=14pc]{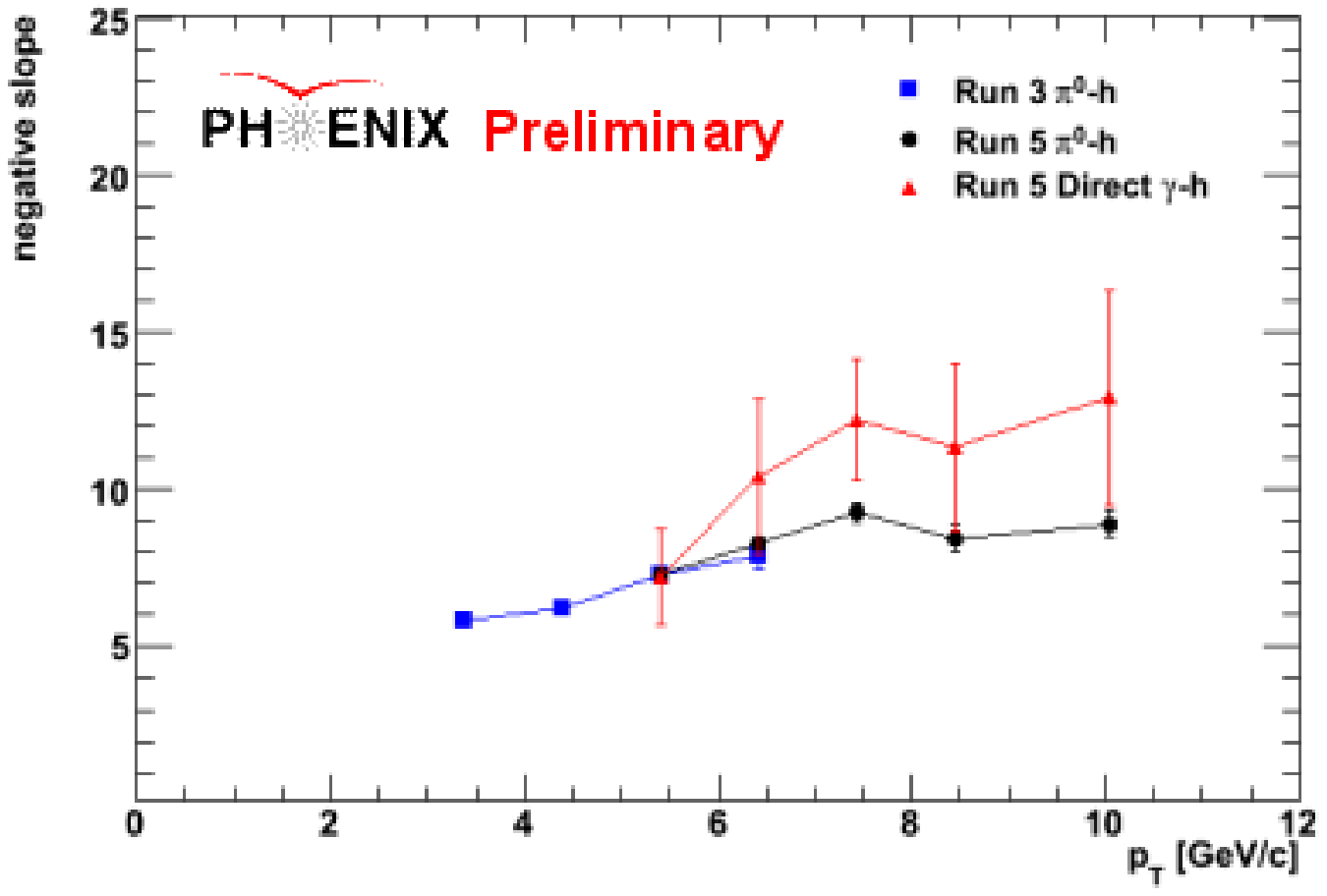} \\ [0.5 cm]
\caption{\label{slope} Slope parameter of exponential fits to \xe\
distributions for \piz\ and direct photon triggers.}
\end{minipage}
\end{center}
\end{figure}

Figure \ref{xe} shows \xe\ distributions for direct photon triggers
from the Run 5 \pp\ data.  The \pu\ range of the charged hadrons is
$1-5$ GeV.  The \xe\ distributions were fit to an exponential in the
range $1.0/p_{T,trigger} < \xe\ < 5.0/p_{T,trigger}$.  The slope
paramter of the \xe\ distributions are shown along with fits to the
Run 3 and Run 5 \piz\ data in figure \ref{slope}.  The \xe\
distributions for direct photon triggers are larger than that of
\piz\ triggers and to be rising with \pu\ within large
uncertainties.  The increased statistics available from the Run 6
data set will better determine whether the direct photon \xe\
distributions scale with \pu.

%\subsection{\AuAu}

\begin{figure}[h]
\begin{center}
\includegraphics[width=16pc]{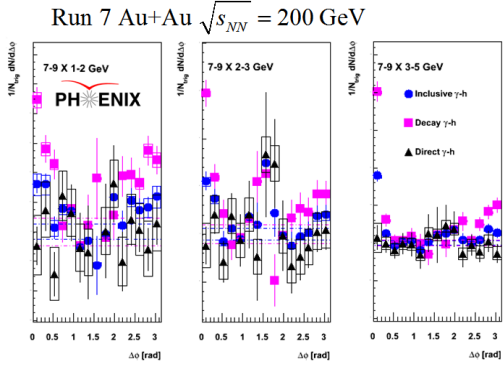}\hspace{2pc}%
\begin{minipage}[b]{14pc}\caption{\label{au} Per-trigger yields of charged hadrons for
inclusive ($\fullcircle$), decay ($\fullsquare$) and direct photon
($\blacktriangle$) triggers vs. $\Delta\phi$ in central \AuAu\
collisions.}
\end{minipage}
\end{center}
\end{figure}

Figure \ref{au} shows per-trigger yields as a function of
$\Delta\phi$ for photons between 7 and 9 GeV and several different
partner \pu\ bins in central \AuAu\ (0-20\%). Inclusive
($\textcolor{blue}{\fullcircle}$), decay
($\textcolor{magenta}{\fullsquare}$) and direct photon
($\blacktriangle$) yields are shown.  The sizeable errors on the
direct photon associated yields correspond to additional
uncertainties in the under-lying event, elliptic flow subtraction
which are not present in the p+p analysis.  The systematic errors
associated with the ZYAM normalization procedure are shown
separately for the inclusive and decay yields.  The ZYAM procedure
is described in \cite{ppg032}.

The near-side yields for direct photon triggers are small enough to
rule out a large contribution from fragmentation.  The away-side
yields are already noticeably suppressed in the inclusive sample and
the suppression in the direct photon sample is
 comparable or greater than for inclusive triggers.

\begin{figure}[h]
\begin{center}
\begin{minipage}{14pc}
\includegraphics[width=14pc]{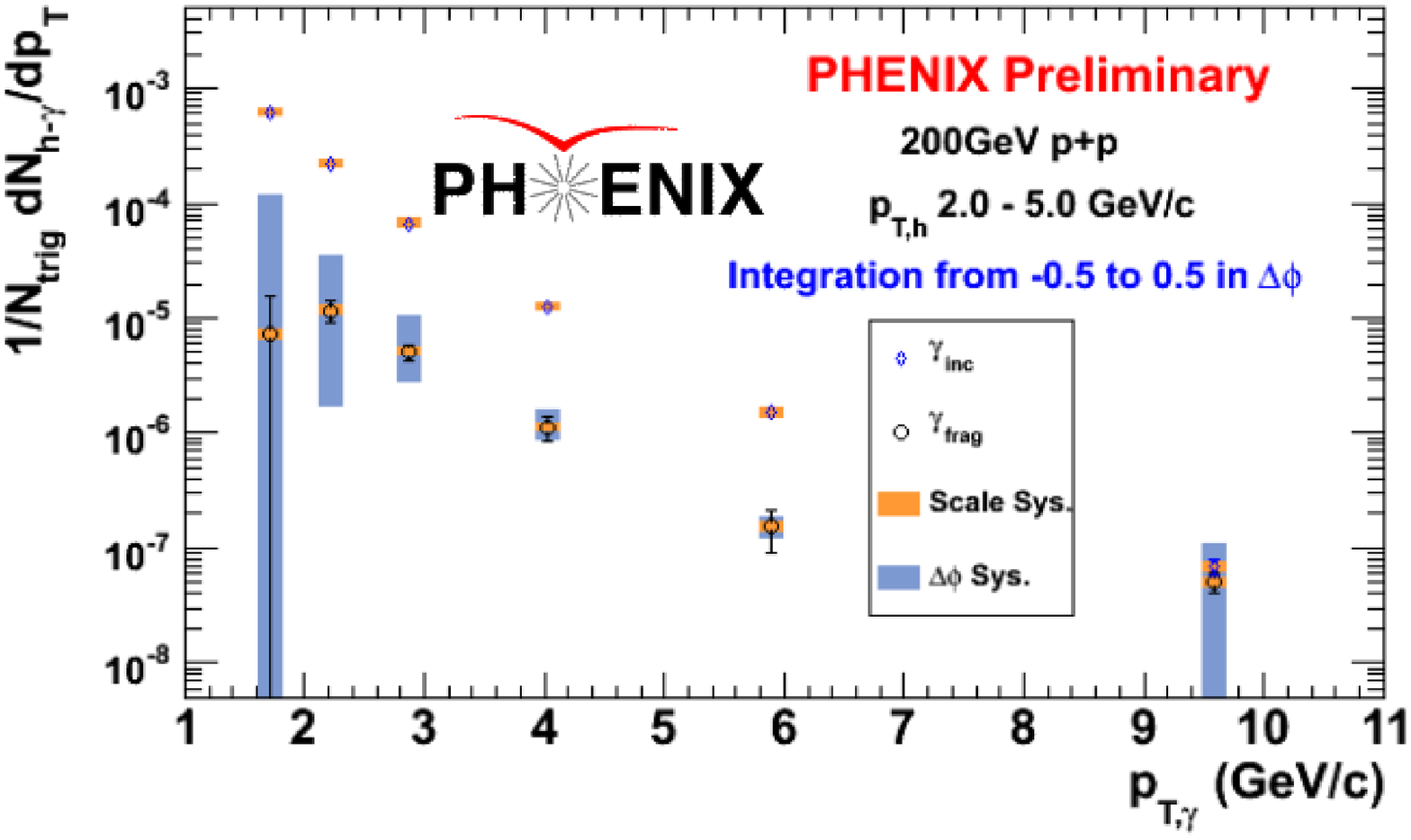} \\ [0.18cm]
\caption{\label{frag2}  Per-trigger yields of associated inclusive
($\opendiamond$) and direct ($\opencircle$) photons as a function of
\pu.}
\end{minipage}\hspace{2pc}%
\begin{minipage}{14pc}
\includegraphics[width=14pc]{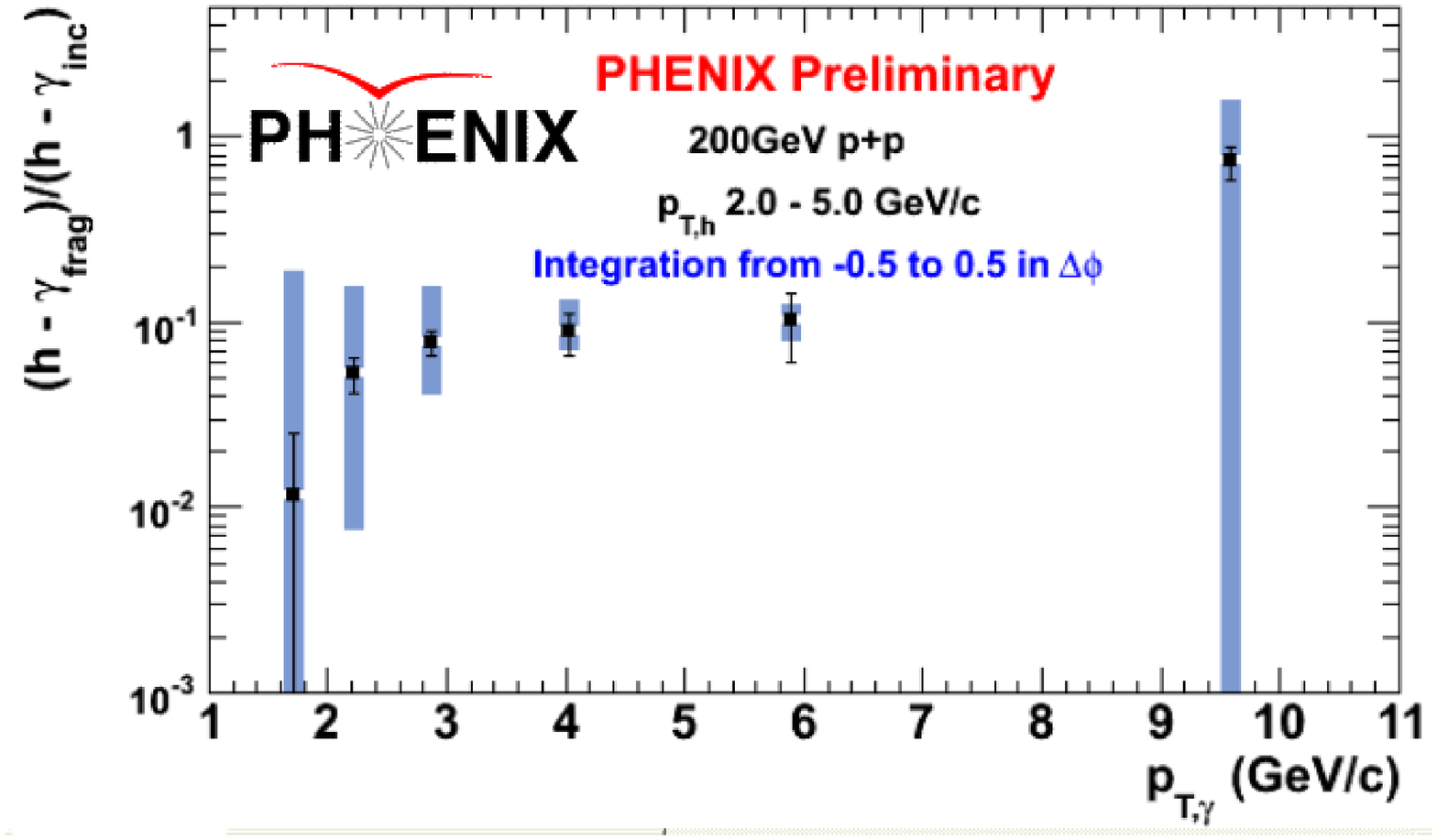}
\caption{\label{frag1} The ratio of the yield of associated direct
photons to associated inclusive photons as a function of photon
\pu.}
\end{minipage}
\end{center}
\end{figure}

The fragmentation component of the direct photon signal should be
measured as a background to the prompt photon component but may also
prove to be an interesting energy loss observable itself. \cite{zak}
By triggering on high \pu\ charged hadrons and looking at the
correlated direct photon production within one radian of $\phi$
($\sim$ the typical near-side jet width) the fragmentation yield can
be analyzed directly. To achieve this, the conditional yield of
inclusive photons, \piz's and $\eta$'s are measured where the \piz\
and $\eta$ are tagged by their invariant mass. The decay yield can
be estimated from the measured \piz\ and $\eta$ by evaluating the
tagging efficiency via a detailed Monte Carlo simulation. By
subtracting the decay contribution from the inclusive sample one
should be left solely with hadron-fragmentation photon pairs since
the prompt photons should seldom be associated with charged hadrons
on the near side. Figure \ref{frag2} shows the per-trigger yield for
charged hadrons in the \pu\ range $2-5$ GeV with inclusive
($\opendiamond$) and direct ($\opencircle$) photons. Figure
\ref{frag1} shows the ratio of the per-trigger yields of
hadron-direct photon pairs to hadron-inclusive photon pairs as a
function of photon \pu.  The uncertainties in the measurement are
smallest for photons between 3 and 7 GeV.  The fraction of
hadron-photon pairs coming from fragmentation in that region is
constrained to be between 5-15 \% after the uncertainties are taken
into account. More work will be done to relate this two particle
measurement to predictions for the rate of single fragmentation
photons.

\section{Conclusions}

Direct Photon-Hadron correlations have been measured in \pp\ and
\AuAu\ collisions using a statistical subtraction method.  Neither
data set shows a large near-side associated yield suggesting that
prompt photons dominate the direct photon sample. For \pp\ the
away-side yield associated with direct photons is similar to the
\piz\ associated yield within errors.  For \AuAu\ collisions the
away-side yield is significantly suppressed.  \xe\ distributions
have been measured from the \pp\ data, but at present the
uncertainties are too large to test whether the \xe\ distribution
scales with \pu. Hadron-Photon correlation measurements have been
performed which measure the yield of hadron-fragmentation photon
pairs directly. For photons in the range 3-7 GeV the fraction of
hadron-photon pairs that come from fragmentation within 0.5 radians
in azimuth is between 5-15 \%.

\end{document}